\documentclass{article}
\usepackage{spconf}
\usepackage{multirow}
\usepackage{amsmath, amssymb}
\usepackage{array}
\usepackage{subfigure}
\usepackage{amsmath}
\usepackage{pseudocode}
\usepackage{multirow}
\usepackage{amsmath}
\usepackage{amssymb}
\usepackage{graphicx}
\usepackage{multirow}
\usepackage{ctable}
\usepackage{caption}
\usepackage{booktabs}
\usepackage{tabulary}
\usepackage{subfigure}
\usepackage{array}
\usepackage{pseudocode}
\begin{document}

\title{Incorporating Betweenness Centrality in Compressive Sensing for Congestion Detection}

\name{Hoda S. Ayatollahi Tabatabaii, Hamid R. Rabiee, Mohammad Hossein Rohban, Mostafa Salehi}
\address{Sharif University of Technology, Tehran, Iran}

\maketitle

\begin{abstract}
This paper presents a new Compressive Sensing (CS) scheme for detecting network congested links. We focus on decreasing the required number of measurements to detect all congested links in the context of network tomography. We have expanded the LASSO objective function by adding a new term corresponding to the prior knowledge based on the relationship between the congested links and the corresponding link Betweenness Centrality (BC). The accuracy of the proposed model is verified by simulations on two real datasets. The results demonstrate that our model outperformed the state-of-the-art CS based method with significant improvements in terms of F-Score. %an average of 11\% and 9\% improvements on F-Score of the retrieved congested links for various number of sparsity, an average of 5\% and 10\% improvements on F-Score for different number of Random Walks (RW) in the network, and an average of 16\% and 15\% on the number of random walks in various sparsity values on two real datasets respectively. It also outperformed the algorithms based on mere betweenness centrality, by an average of 12\% on F-Score for different sparsities and 69\% on F-Score for various random walk numbers.
\end{abstract}

\section{Introduction}
The network tomography scheme is a method to calculate link variables such as delay or bandwidth, using the end-to-end measurements. In this paper, we intend to solve a congestion detection problem using the network tomography scheme, based on end-to-end measurements. Our goal is to decrease the number of required measurements in order to identify all congested links. Since only a small number of network links may become congested, it is assumed that the delay vector (which contains delays of all links) is sparse. %That is, most elements of this vector are zero and only the elements related to the congested links has non-zero values. 

To solve this problem, we employ Compressive Sensing (CS) \cite{CSIntro} which has received significant attention in the recent years. Since its inception, CS shows remarkable results in reconstructing a high dimensional sparse signal with a small set of measurements. A significant issue in this field is the sample complexity which is the number of required measurements for high quality signal recovery which is solvable through LASSO \cite{CSIntro}. Although CS is a new sampling scheme in signal processing, it has been applied to many network applications specially in the area of network tomography. Fault diagnosis \cite{NetTom2}, traffic estimation \cite{NetTom1}, data localization \cite{CSP2P}, and congestion detection \cite{CSoverGraph} are some well known applications in this field.

The state-of-the-art CS-based algorithm in network tomography \cite{CSoverGraph}, used random walks to gather end-to-end delays and employed LASSO in its model for congestion detection.  However, it has two major drawbacks; relatively high number of random walk measurements (much higher than the total network links), and low accuracy in detecting the congested links when the required number of measurements is small. 
In this paper, we try to solve those drawbacks by using the prior knowledge corresponding to the correlation of link betweenness centrality (the number of shortest paths that traverse from the link) and link congestion. We also introduce a new CS objective function in which we consider the dependency of congestion and betweenness centrality. We further show that the proposed objective function can be considered as the Elastic-Net model \cite{ENet}, which is a more stable alternative to LASSO.
We have simulated the proposed Compressive Sensing in Congestion Detection (CSCD) model, by using real data with various configurations. %in terms of sparsity of congested links and the number of measurements (random walks) in the network. %The results demonstrate a significant improvement in detecting the congested links, compared to the method presented in \cite{CSoverGraph}. Specifically, w
We have compared the F-Score (of the retrieved congested links) in our model with \cite{CSoverGraph}, in terms of the number of measurements (random walks) and various sparsity of the network delays in which we achieve significant improvements. 

The rest of the paper is organized as follow; in Section \ref{Related Work}, we describe the related work in network tomography, congestion detection, and compressive sensing. Section \ref{Problem Definition} shows an introduction to compressive sensing and the problem we try to solve in this paper. In Section \ref{The Proposed Model}, we introduce the proposed compressive sensing model. Comparison with the previous model is appeared in Section \ref{Evaluation}. Finally, we conclude the paper in Section \ref{Conclusion}. 
%%---------------------------------------------------Related Work-----------------------------------------------
\section{Related Work}
\label{Related Work}
%Nyquist sampling is a well-known signal compression scheme in signal processing for the band-limited signals. In CS, however, the main goal is to compress sparse signals which are not band-limited. Since compressive sensing can significantly reduce the number of required measurements to recover the sparse signals, it is an alternative for sparse signal compression. CS can be applicable for compression and recovering the required information in both peer-to-peer and wireless networks. \cite{CSP2P} addressed the problem of peers' accessibility to the network global information. Each peer is given a local access to the global sparse information in the network. Employing an integration of random walk and compressive sensing, the gathered information is compressed and distributed by decreasing the communication overhead. The problem of gathering distributed data in wireless sensor networks is discussed in \cite{routing} using multi-hop routing and compressive sensing. In \cite{CWS}, a random projection of the sensed data in a wireless network is sent to a Fusion Center (FC) in order to reconstruct the data with a good approximation and also reconstruct the data with the least total power, distortion, and least latency required for data transfer to the FC. 
%Moreover, in 
Compressive sensing in network tomography is first discussed in \cite{GroupTest1}. The authors discussed a group testing problem on the Erd$\acute{\text{o}}$s-R$\acute{\text{e}}$nyi random graphs. They have applied OR operation (instead of summation) on the gathered measurements and calculated the required number of measurements when the link variables are binary. %, the number of required measurements to recover a sparse vector is in the order of $O(k^2 \log{\frac{N}{k}})$ .
%In network tomography \cite{NetTomography}, compressive sensing can be used efficiently for signal reconstruction. The aim of network tomography is to recover some network links or nodes characteristics using the end-to-end measurements. 
In \cite{CSoverGraph}, the authors gathered end-to-end delay information by applying random walks to the network and recovered the sparse delay values of network links using compressive sensing. Although they have better theoretical order for the number of measurements compared to \cite{GroupTest1}, they have used high number of random walks to achieve good recovery percentage in practice. In their theoretical proof, they required that the graph is highly connected which may not be true in all of the real networks. Moreover, their practical results are accomplished according to specific graphs such as the complete graph.

In our model, however, we try to solve these practical issues by employing the relationship between link betweenness centrality and link congestion in the network. As noted in \cite{BC1}, link betweenness centrality and congestion are two measurements in the network which are highly correlated. In  \cite{BC1}, \cite{BC2}, and \cite{BC3} link betweenness centrality is used as the only element to detect the congested links in the network. Although betweenness centrality is effective to achieve this purpose, none of these papers consider end-to-end measurements in finding the congested links. Later in this paper, we show  how efficient it is to employ both link betweenness centrality and end-to-end measurement for congestion detection.

%In another network tomography problem, by using expander graphs \cite{NetTomographyviaCS}, the network routing matrix is designed in a way that the minimum number of measurements is achieved for recovering links' delays. In \cite{NetMonitor1} and \cite{NetMonitor2}, diffusion wavelets are designed based on routing matrix in order to find a basis in which the signals are compressible. In the integration of network tomography and expander graphs, \cite{ExpanderGraph1} applies some conditions on the routing matrix in order to estimate link delays of the network from  end-to-end measurements.
%In the networking applications, useful prior knowledge can be defined to model the desired signals. Recently, model-based compressive sensing \cite{ModelCS} is proposed a framework to use a signal model (other than signal sparsity) for reducing the sample complexity. In contrast to the previous CS methods in the network, we are going to employ prior knowledge in the model-based CS to decrease the required number of measurements for signal recovery. 
%---------------------------------------------------Problem Definition-----------------------------------------------
%\parskip -3pt
\section{Problem Definition}
\label{Problem Definition}
We consider a real communication network as an undirected graph with the set of links $E$, and vertices $V$. We assume each link can transfer data in both directions between two connected nodes. We are going to measure the network congested links using end-to-end delays of $M$ random walk paths in the network ($M<|E|$). Thus, we should first measure the delay of each network link. Let the vector $\mathbf y_{M \times 1}$ denote the observed end-to-end delays where $\mathbf y_i$ represents the end-to-end delay of the path $i$. We also define $\mathbf x_{N \times 1}$ as the links' delay vector where $\mathbf x_j$ represents the delay of link $j$. Since the number of congested links in a network are sparse, $\mathbf x$ is considered as a $k$-sparse vector ($k \ll N$ and $N=|E|$). The goal is to recover vector $\mathbf x_{N \times 1}$ which contains the delay of all $N$ links of the network. Moreover, by recovering $\mathbf x$, the congested links can be recognized through their high amount of delay. To recover this vector, compressive sensing method has been used. The recovery process starts with solving the following equation: 
\begin{equation}
\label{y_Ax}
\mathbf y = A \mathbf x + \epsilon
\end{equation}
where $\epsilon$ is the noise vector that is experienced in the random walk measurements, and $A$ is an $M \times N$ matrix. The elements of $i^{th}$ row and $j^{th}$ column of $A$ is equal to $1$, if the $i^{th}$ link of the network is available in the $j^{th}$ random walk path. We employ $M$ random walks over the network graph and gather information regarding the end-to-end delay of each path ($\mathbf y_i$). Having $A$ and $\mathbf y$, we may reconstruct $\mathbf x$ by minimizing $\lVert\mathbf x\rVert_{0}$
% $\ell^0\text{norm}(\mathbf x)$ 
under the constraint of Eq. \eqref{y_Ax}. However, as stated in \cite{l0norm}, this is an NP-hard optimization problem. In compressive sensing, the aforementioned optimization problem is solved by minimizing the 
%$\ell^1\text{norm}(\mathbf x)$ 
$\lVert\mathbf x\rVert_{1}$ instead of $\lVert\mathbf x\rVert_{0}$
%$\ell^0\text{norm}(\mathbf x)$ 
which results in an optimization problem with computational complexity of $O(N^3)$ \cite{l0norm}, known as LASSO: 
\begin{equation} 
\label{CSEq}
\hat{\mathbf x} = \arg\min\limits_{\mathbf x}~~ (\lambda \lVert \mathbf x \rVert_{1} + \gamma \lVert A \mathbf x - \mathbf y \rVert^{2}_{2})
\end{equation}
%------------------------------------------Proposed Model---------------------------------------------------------
\section{The Proposed Model}
\label{The Proposed Model}
We have employed the Bayesian theory in our model which is used in many compressive sensing applications \cite{Bayesian1} \cite{Bayesian2} \cite{Bayesian3}. The previous applications, however, have not considered prior knowledge on $\mathbf x$ other than its sparsity. The main goal of our model is congestion detection with a few number of random walk measurements. We have employed the betweenness centrality \cite{BetweenC} of the congested links as the prior knowledge in $\ell^1\text{-norm}$ minimization. 

Let $D$ denote the maximum delay that each link can tolerate in the network and $b_i$, $0 \leq b_i \leq B$, denote the link betweenness of the link $i$ where $B$ is the maximum link betweenness centrality in the network. We use linear interpolation to capture the relation between the betweenness centrality and the prior belief about the delay values of network links. In this way, we get the scaled version of link betweenness centralities ($\mathbf s$). 

With a high probability, a link with the maximum link betweenness ($B$) has the highest delay ($D$) in the network \cite{CongBetween}. Moreover, a link with the lowest link betweenness (0) is more likely to have no delay. Thus, (0,0) and ($B$,$D$) are two points on the interpolating line. Therefore, by considering the link betweenness values as the X-axis ($b_{i}$) and our prior belief about links' delay as the Y-axis ($s_{i}$) in a two dimensional space, $s_{i}$ is given by: 
\begin{equation}
 s_{i} = D \times \frac{b_i} {B}%{\text{Max}_{j=1}^{N} b_j }
\end{equation}
%Prior density function of $\mathbf x$ should be defined in a way that both the sparsity of $\mathbf  x$, and the correlation of delays in $\mathbf x$ and betweenness centrality are taken into account. 
In order to recover $\mathbf x$ from $\mathbf y$, it is critical that $\mathbf x$ be sparse. Since $\mathbf s$ is also a sparse vector (because it is highly correlated with $\mathbf x$), $\mathbf x - \mathbf s$ should be sparse too. Therefore, we define the probability density function of $\mathbf x$ as follows:
%\vspace{-3mm}
\begin{equation}
\label{ProbX}
P(\mathbf x) ~\propto ~\exp ~ \left\{ -\left( \frac{\lVert\mathbf x - \mathbf s\rVert_1}{k_1} + \frac{\lVert\mathbf x-\mathbf s\rVert^2_2}{k_2} \right)\right\}
\end{equation}
where $k_1$,$k_2 \in \mathbb{R}^\text{+}$. It penalizes non-sparse choices of $\mathbf x - \mathbf s$ (by the first term) and also vectors $\mathbf x$ which are not similar to $\mathbf s$ (by the second term).
By observing $\mathbf y$, we intend to find $\mathbf x$ with the highest probability. Thus, we may maximize $P(\mathbf x|\mathbf y)$ by using the Maximum a Posteriori probability (MAP) estimation as follows:
\begin{equation}
\label{PX_Y}
\max\limits_{\mathbf x}~~\left(P(\mathbf x|\mathbf y) = \frac{P(\mathbf y|\mathbf x)~P(\mathbf x)}{P(\mathbf y)}\right)
\end{equation}
Since the goal of Eq. \eqref{PX_Y} is to find the maximum value of $\mathbf x$ in $P(\mathbf x|\mathbf y)$, by eliminating the terms that do not depend on $\mathbf x$, and taking logarithm on both sides, we obtain:
\begin{equation}
\label{LogP_X}
\begin{split}
&\max\limits_{\mathbf x}~\log \left( P(\mathbf y|\mathbf x)~ P(\mathbf x) \right) =
\max\limits_{\mathbf x}~\left( \log P(\mathbf y|\mathbf x)+\log P(\mathbf x) \right)
\end{split}
\end{equation}
On the other hand, by observing $\mathbf y$, we intend to find $\mathbf x$ from Eq. \eqref{y_Ax}. We assume $\epsilon$ has a normal distribution with zero mean and covariance matrix $\sigma^2I$. Therefore:   
\begin{equation}
\label{PY_X}
P(\mathbf y|\mathbf x) \sim \mathcal{N}(A\mathbf x,\sigma^2 I)
\end{equation}
According to Eq.s \eqref{ProbX} and \eqref{PY_X}, Eq. \eqref{LogP_X} may be expanded 
as follows:
\begin{align*}
&\max\limits_{\mathbf x}~\left( \log P(\mathbf y|\mathbf x)+\log P(\mathbf x)\right) \equiv \\
&\min\limits_{\mathbf x}~ \Bigg(-\log \left\{ \frac{\exp \left( -(\mathbf y-A\mathbf x)^{\text{T}} \frac{1}{2\sigma^2} I (\mathbf y-A\mathbf x)\right)}{(2\pi)^{\frac{M}{2}} \text{det}(\sigma^2 I)^\frac{1}{2}} \right\} - \\
&~\quad\quad\quad\quad\log ~ \left\{ \exp -\left( \frac{\lVert\mathbf x - \mathbf s\rVert_1}{k_1} + \frac{\lVert\mathbf x-\mathbf s\rVert_2^2}{k_2}\right) \right\} + C \Bigg) 
\end{align*}
where $C \in \mathbb{R}$ and $M$ is the number of random walk measurements. Therefore, the optimization problem becomes:
%\begin{equation}
%\begin{split}
%\min\limits_{\mathbf x}~~~ \left( \gamma\lVertA\mathbf x-\mathbf y\rVert_2^2 + \lambda\lVert\mathbf x - \mathbf s\rVert_1 + \alpha \lVert\mathbf x - \mathbf s\rVert_2^2 \right)
%\end{split}
%\end{equation}
 %Thus, the network delay vector ${\mathbf x}$ can be recovered as follows:
\begin{equation}
\label{OurCSEq}
\hat{\mathbf x} = \arg\min\limits_{\mathbf x}~~\left( \lambda \lVert \mathbf x - \mathbf s \rVert_{1} + \gamma \lVert A \mathbf x - \mathbf y \rVert^{2}_{2} + \alpha \lVert \mathbf x - \mathbf s \rVert_{2}^{2} \right)
\end{equation} 
where $\gamma = \frac{1}{\sigma^2}$, $\lambda = \frac{1}{k_{1}}$, and $\alpha = \frac{1}{k_{2}}$.
An important goal in CS is to have model consistency \cite{ENet}, which shows that the support of recovered $\mathbf x$ converges to the support of original $\mathbf x$ as the number of random walk measurements goes to $\infty$. To verify this property, we show that our problem is similar to the one discussed in \cite{ENet} which is known as Elastic-Net:
\begin{equation} 
\hat{\mathbf \beta} = \arg\min\limits_{\mathbf \beta}~~\left( \lambda \lVert \mathbf \beta \rVert_{1} + \gamma\lVert A\mathbf \beta - \mathbf y^{\prime}\rVert^{2}_{2} + \alpha \lVert \mathbf \beta \rVert_2^2 \right)
\label{ENetEq}
\end{equation} 
To increase the recovery accuracy, achieve model consistency, and overcome LASSO limitations \cite{ENet}, Elastic-Net is used as an alternative to LASSO.

Consider $\beta = \mathbf x - \mathbf s$ and $\mathbf y^{\prime} = A\mathbf s - \mathbf y$. Since $\mathbf s$ is a constant matrix, it is easy to show that Eq. \eqref{OurCSEq} and Eq. \eqref{ENetEq} are two equivalent optimizations. Thus, our model is in the form of Elastic-Net.
%Assume $\mathbf x - \mathbf s = \mathbf z$, then $A \mathbf x - \mathbf y = A (\mathbf z + \mathbf s) - \mathbf y$. Since $\mathbf s$ is a constant matrix, we may consider $A\mathbf s - \mathbf y = \mathbf y^{\prime}$. Thus, $A \mathbf x - \mathbf y = A \mathbf z - \mathbf y^{\prime} $. Moreover, since $\mathbf s$ contains constant values,$\arg\min\limits_{\mathbf x - \mathbf s} = \arg\min\limits_{\mathbf x}$ and ${\hat\mathbf x - \mathbf s} = \hat{\mathbf x}$. Therefore, by having  $\beta = \mathbf x - \mathbf s$, our model is in the form of Elastic-Net.

As mentioned before, we have $M$ random walk measurements, $N$ network links and $k$ is the sparsity value. 
Without loss of generality, assume that only the first $k$ elements of $\mathbf x$ are non-zero. Then $\mathbf x_{1} = (x_{1}, ... ,x_{k})$, $\mathbf x_{2} = (x_{k+1}, ... ,x_{N})$, $A_{(1)}$ contains the first $k$ columns of $A$, and $A_{(2)}$ contains the last $N-k$ columns of $A$.

Therefore, by having $C_{11} = \frac{1}{M}A_{(1)}^{\text{T}} A_{(1)}$, $C_{12} = \frac{1}{M}A_{(1)}^{\text{T}} A_{(2)}$, $C_{22} = \frac{1}{M}A_{(2)}^{\text{T}} A_{(2)}$, and $C_{21} = \frac{1}{M}A_{(2)}^{\text{T}} A_{(1)}$, Irrepresentable Condition (IC) can be shown to be a necessary and sufficient condition for LASSO's model consistency \cite{ENet}:
\begin{equation} 
\label{IC_eq}
\exists~\eta>0 : \Bigg\lVert C_{\text{21}} (C_{\text{11}})^{-1}(\textit{sign}(\mathbf x_{1})) \Bigg\rVert_{\infty} \leq 1 - \eta
\end{equation} 
Moreover, according to the Corollary 1 in \cite{ENet}, if the Elastic Irrepresentable Condition (EIC) is satisfied, the Elastic-Net model has model consistency by choosing right values for $\lambda$, $M$, and $\alpha$. The EIC is as follows:
\begin{equation} 
\label{EIC_eq}
\Bigg\lVert C_{\text{21}} (C_{\text{11}} + \frac{\alpha}{M} \mathbf I)^{-1}\Bigg(\textit{sign}(\mathbf x_{1})+\frac{2\alpha}{\lambda}\mathbf x_{1} \Bigg) \Bigg\rVert_{\infty} \leq 1 - \eta
\end{equation} 
where $\eta > 0$.
In the next section, we show that in our model EIC is satisfied with a higher probability compared to IC.
 %--------------------------------------------Evaluation-------------------------------------------------------------
\section{Evaluation}
\label{Evaluation}
\subsection{Simulation Framework}
In order to evaluate the proposed model, we have performed extensive simulations (30 runs) in MATLAB. We have used two real datasets. The first dataset contains the information of a mobile operator which has 273 links and 158 nodes corresponding to the network devices in Mobile Switching Centers (MSC) placed in 40 cities. The second dataset contains the information of a data network with 366 links and 277 nodes which are located in more than 50 cities. We have also considered the variance of the results by measuring the related error bars. For the first dataset, we consider  $\lambda = 10^{-3}$, $\alpha = 10^{-5}$, and $\gamma = 1$. For the second dataset we have  $\lambda = \alpha = \gamma = 1$. Through several simulations, these are almost the best configurations for both LASSO and our model. The number of steps in each random walk is assumed to be 15.
%---------------------------------------------------Validation----------------------------------------------
\subsection{Validation}
\label{Validation}
To evaluate the proposed model, named Compressive Sensing in Congestion Detection (CSCD), we have used $\text{F-Score}$ measure which corresponds to the harmonic mean of precision and recall. Precision measures the percentage of correctly detected congested links to the sum of correctly and wrongly detected congested links, and recall refers to the percentage of correctly detected congested links to the total detected congested links. 

First, we have evaluated the proposed CSCD model and LASSO through model consistency. Assuming $\eta = 0.01$, EIC in Eq. \eqref{EIC_eq} and IC in Eq. \eqref{EIC_eq} are verified for our model and LASSO, respectively. At the end of all 30 simulation runs, the percentage of the times that EIC and IC are satisfied, is computed.
\begin{figure}[h]
\centering
\includegraphics[width=0.27\textwidth]{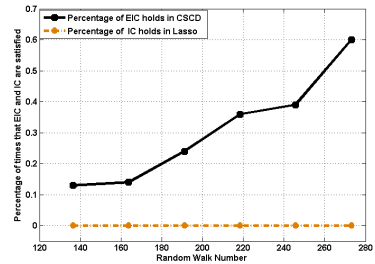}~ 
\caption{Comparison of probability that EIC holds in Elastic-Net and IC holds in LASSO in various number of random walks in the first dataset}
\label{ENet-LASSO}
\end{figure}
It has to be mentioned that CS-over-Graphs in \cite{CSoverGraph} has used LASSO in its model. Moreover, in the theoretical proof of \cite{CSoverGraph}, they required that the graph is highly connected which may not be true in all of the real networks. Choosing $\lambda = \sqrt{M} \log M$, $k = 8$ as sparsity, and $N = 273$ as the number of links in the first dataset, by increasing $M$ such that $M \rightarrow \infty$, Fig.\ref{ENet-LASSO} shows that in CSCD model model consistency holds with higher probability.

\begin{figure}[h]
\centering
\includegraphics[width=0.28\textwidth]{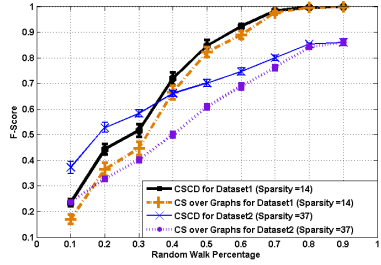}
\caption{F-Score versus random walk number in two datasets}
\label{F-ScoreRW}
\end{figure}

\begin{figure}[h]
\centering
\includegraphics[width=0.28\textwidth]{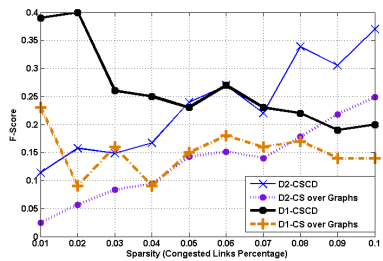} 
\caption{F-Score versus sparsity in two datasets}
\label{F-ScoreSparsity}
\end{figure}

\begin{figure}[h]
\centering
\includegraphics[width=0.28\textwidth]{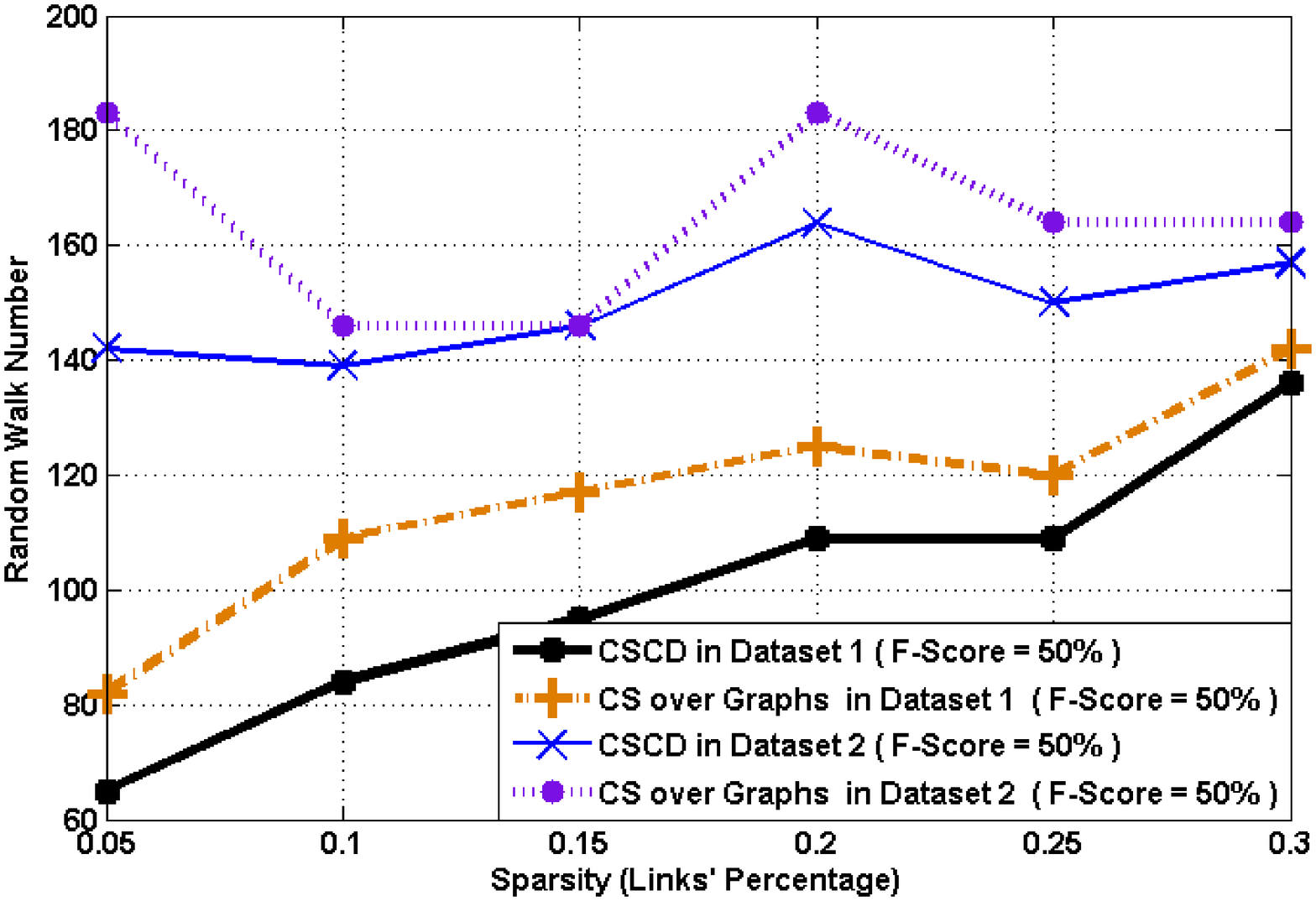}~ 
\caption{Sparsity versus various number of random walks in two datasets }
\label{RWSparsity}
\end{figure}

\begin{figure}[h]
\centering
\subfigure[]{\label{RW-BC}\includegraphics[width=0.27\textwidth]{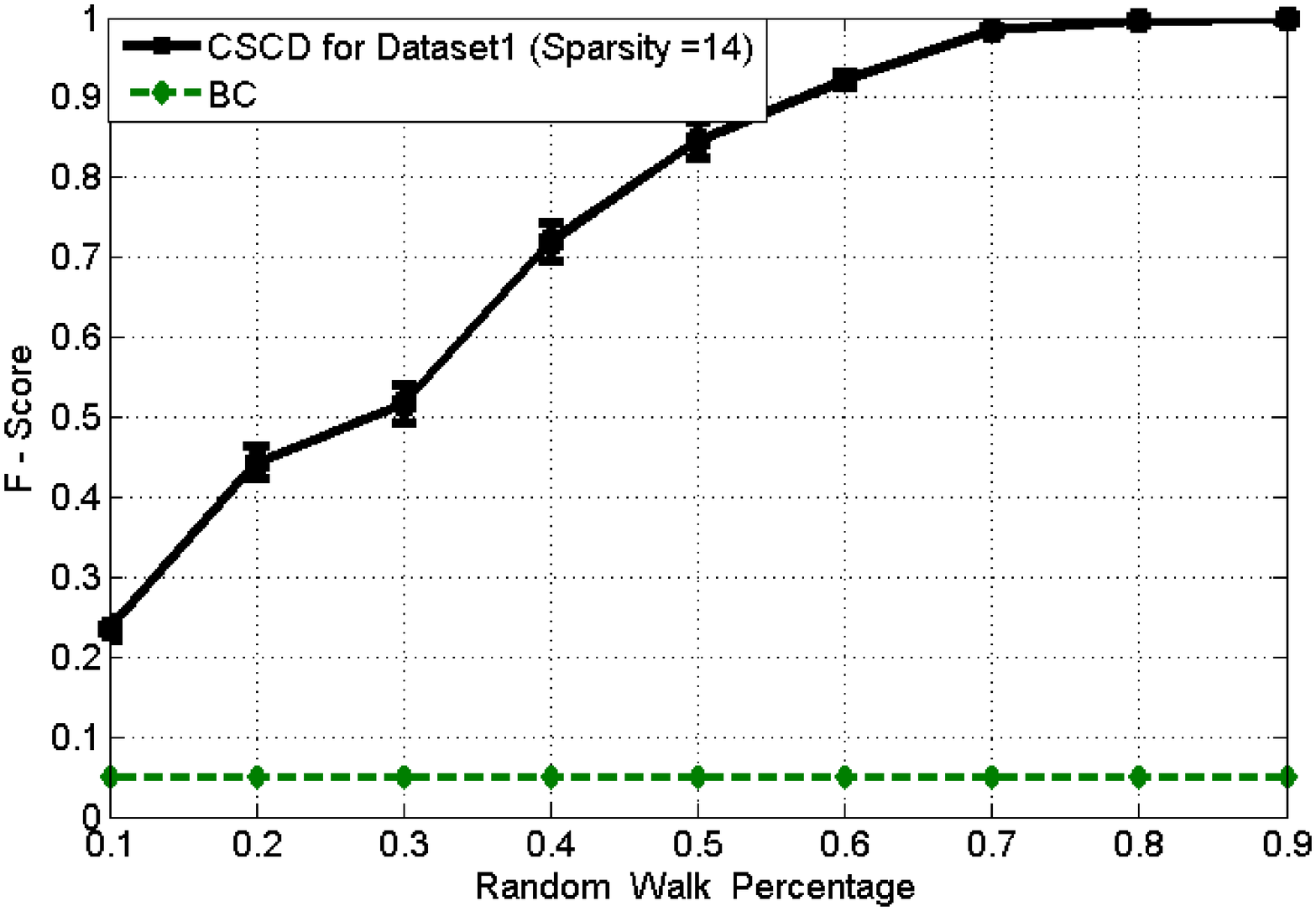}}~
\subfigure[]{\label{Sparsity-BC}\includegraphics[width=0.27\textwidth]{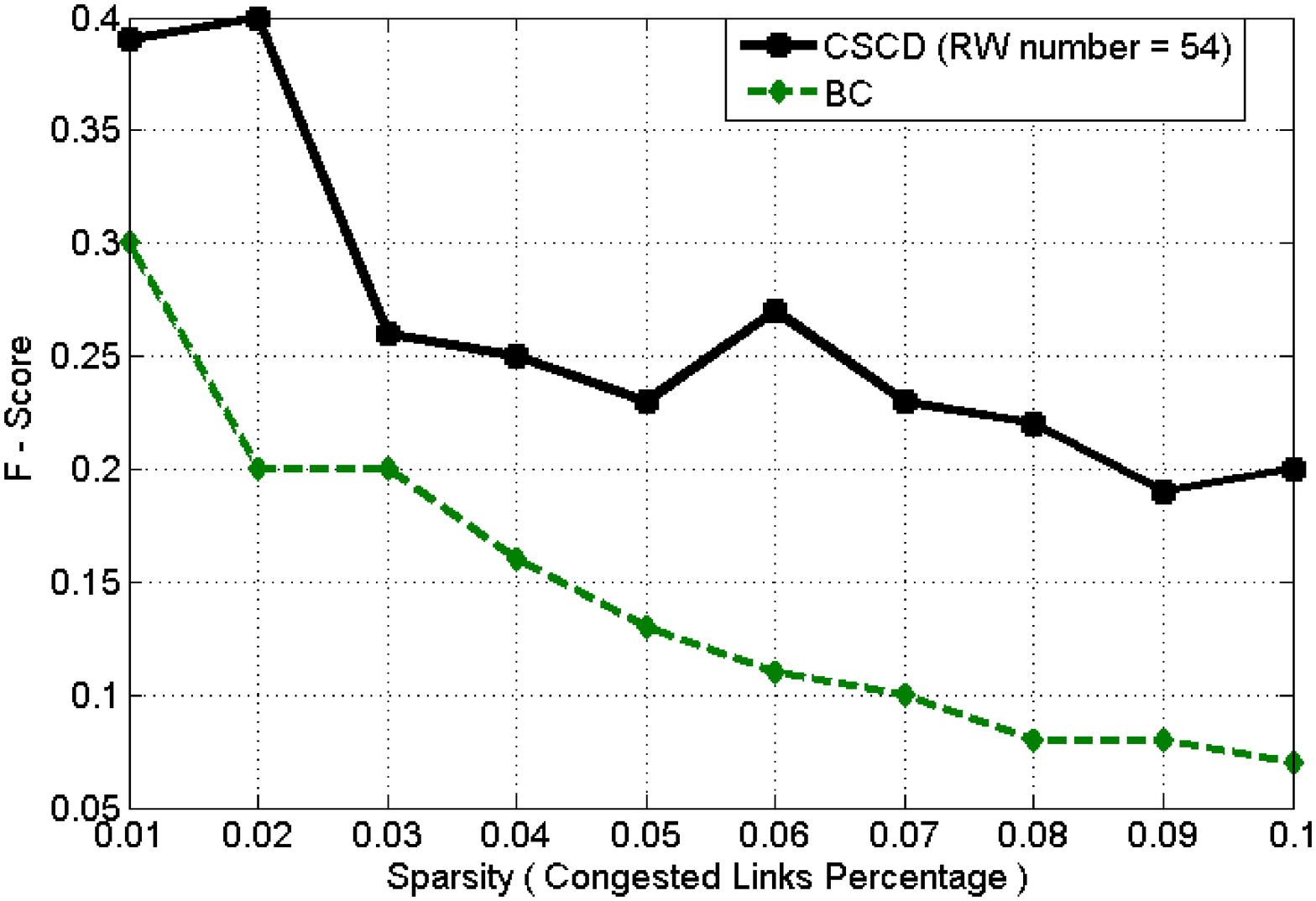}}
\caption{Comparison of our model with BC algorithm in terms of (a) the required number of random walks (b) sparsity}
\label{MereBC}
\end{figure}

The rest of this section shows the evaluation of F-Score in our model in various settings. Fig. \ref{F-ScoreRW} illustrates an improvement of the F-Score performance of the CSCD model by an average of 5\% (Dataset 1) and 10\% (Dataset 2) for the various number of random walks compared to the Compressive Sensing over Graphs method (CS-over-Graph) presented in \cite{CSoverGraph}. The number of random walks changes from 10\% to 90\% of the total network links. Although, we have evaluated F-Score for various random walk steps, the results were similar to those shown in Fig. \ref{F-ScoreRW}. 
Clearly, for lower number of random walks, we have lower number of measurements, and thus less samples. Since the proposed model simultaneously employs the prior knowledge based on the link betweenness centrality, it performs better than CS-over-Graphs model \cite{CSoverGraph} where only the sparsity information ($\lVert\mathbf x\rVert_{0}$) is used. However, as the number of random walks grows, our F-Score gets closer to CS-over-Graphs'. Because at higher number of random walk the prior knowledge becomes less significant. %Moreover, we have improved F-Score by 12\% compared to \cite{CSoverGraph}.
As illustrated in Fig. \ref{F-ScoreSparsity}, we have also evaluated the F-Score of CSCD in terms of sparsity of the congested links in the network ($\lVert {\mathbf x}\rVert_0$). The F-Score of the CSCD model is improved by an average of 11\% in the first dataset, and 9\% in the second one compared to CS-over-Graphs model.
In Fig. \ref{RWSparsity}, we have measured the required number of random walks in terms of the sparsity of the congested links in the network ($\lVert\mathbf x\rVert_0$). Considering the sparsity varies from 5\% to 30\% of the network links, we have calculated the least required number of random walks when F-Score equals to 50\%. The number of random walks in CSCD is decreased by an average of 16\% in the first dataset and 15\% in the second one compared to CS-over-Graphs. %Since in CSCD, we employ a prior knowledge as long as the sparsity of $\mathbf x$, the required number of random walks to detect the congested links is reduced compared to \cite{CSoverGraph}.
%\subsection{Comparison with mere BC algorithm}
%In this section 
We also compare the F-Score of our model with the algorithm that is used only Betweenness Centrality (BC) measurement for congestion detection (BC algorithm) as employed in \cite{BC1}, \cite{BC2}, and \cite{BC3}. The result is illustrated in Fig. \ref{MereBC}.
%In Fig.s \ref{Sparsity-BC} and \ref{RW-BC}, we measure F-Score of CSCD model and compare the results with the BC algorithm in terms of sparsity and RW numbers. %In \ref{Sparsity-BC}, we have changed the sparsity of congested links from 1\% to 10\% of network links and evaluate CSCD model RW step of 15 in the first dataset.
%\vspace{-3mm}
Since betweenness centrality is independent from the number of random walks, it remains constant in Fig. \ref{RW-BC}.
As shown in Fig. \ref{MereBC}, CSCD outperformed the algorithms based on mere betweenness centrality, by an average of 12\% on F-Score for different sparsities and 69\% on F-Score for various random walk numbers.
%As illustrated in Fig. \ref{RW-BC}, the F-Score of BC algorithm stays in the constant value of 7\% while the F-Score of CSCD model grows as the number of RW increases. The main reason of this event is that the betweenness centrality is measured based on the number of shortest paths that traverse through the network links. Thus, it is independent from the number of random walks.
%---------------------------------------------Conclusion---------------------------------------------------
%\vspace{-3mm}
\section{Conclusion} 
\label{Conclusion}
In this paper, we introduced a new objective function based on the concepts of compressive sensing in a network tomography application. We used the link betweenness prior knowledge in our objective function which results in a decrease on the required number of measurements for detecting the network congested links. Based on extensive simulation results, we verified significant improvement in accuracy of detecting the network congested links in two real datasets. %The F-Score improvement of CSCD model is 11\% and 9\% for different values of sparsity and  5\% and 10\% for various number of random walks compared to the state-of-the-art CS-over-Graphs model in \cite{CSoverGraph}. We also decrease the number of required RW to reach the F-Score of 50\% by an average of 16\% and 15\% on two real datasets respectively. Moreover, we achieved an average of 12\% improvement on F-Score in terms of sparsity and 69\% on F-Score in terms of RW numbers compared with the BC algorithm.
\bibliography{mybib}
\bibliographystyle{IEEEbib}
\end{document}